\documentclass{article}
\usepackage{times}
\usepackage{amsfonts}
\usepackage{graphicx}
\usepackage[pdfmark]{hyperref}
\begin{document}
\noindent
{\Large  A GAUGE THEORY OF QUANTUM MECHANICS}
\vskip1cm
\noindent
{\bf Jos\'e M. Isidro}${}^{1,2}$ and {\bf Maurice A. de Gosson}${}^{3}$\\
${}^{1}$Instituto de F\'{\i}sica Corpuscular (CSIC--UVEG), Apartado de Correos 22085,\\ Valencia 46071, Spain\\
${}^{2}$Max--Planck--Institut f\"ur Gravitationsphysik, Albert--Einstein--Institut,\\ D--14476 Golm, Germany\\
${}^{3}$Institut f\"ur Mathematik, Universit\"at Potsdam, Am Neuen Palais 10,\\ D--14415 Potsdam, Germany\\
{\tt jmisidro@ific.uv.es},  {\tt maurice.degosson@gmail.com}\\

\noindent
{\bf Abstract} An Abelian gerbe is constructed over classical phase space. The 2--cocycles defining the gerbe are given by Feynman path integrals whose integrands contain the exponential of the Poincar\'e--Cartan form.  The U(1) gauge group on the gerbe has a natural interpretation as the invariance group of the Schroedinger equation on phase space.


\section{Introduction}\label{kinft}

Quantum mechanics on phase space, pioneered by Wigner \cite{WIGNER} in 1932, has received renewed attention recently \cite{GOSSON, ZACHOS}. In this paper we relate the approach to phase--space quantum mechanics presented in refs.  \cite{LETTER, MAURICE} with the approach to quantisation via gerbes \cite{GERBES} introduced in ref. \cite{PHASEGERBES}. Our conclusions can be summarised in the statement that {\it symplectic covariance of the Schr\"odinger equation on phase space}, in the sense of refs. \cite{LETTER, MAURICE},  {\it is equivalent to gauge invariance under a U(1) gerbe on phase space}, the latter invariance understood as in ref. \cite{PHASEGERBES}. Our results thus lead to a gauge theory of quantum mechanics on phase space. However this gauge theory is not of the usual Yang--Mills type (a potential 1--form and a field--strength 2--form). Rather, gauge invariance here is in the sense of U(1) gerbes with a connection \cite{GERBES}: a potential 1--form $A$, a potential 2--form $B$ (or {\it Neveu--Schwarz} field) and a field strength 3--form $H$.

\section{The gerbe}\label{rabon}

In this section we summarise the results of ref. \cite{PHASEGERBES} concerning the construction of an Abelian gerbe with a connection on a $2d$--dimensional phase space $\mathbb{P}$. Let a mechanical action 
\begin{equation}
S:=\int_{\mathbb{I}}{\rm d}t\,L
\label{rmlcrlzl}
\end{equation}
be given as the integral of the Lagrangian $L$ over a certain time interval $\mathbb{I}\subset\mathbb{R}$. On the open set $U_{\alpha}\subset\mathbb{P}$ we can pick Darboux coordinates $q^j_{(\alpha)}, p_j^{(\alpha)}$ such that the restriction $\omega\vert_{U_\alpha}$ reads
\begin{equation}
\omega=\sum_{j=1}^d{\rm d}q^j\wedge{\rm d}p_j,
\label{bienn}
\end{equation}
where we have dropped the index $\alpha$. The canonical 1--form $\theta$ on $\mathbb{P}$ defined as \cite{MS}
\begin{equation}
\theta:=-\sum_{j=1}^dp_j{\rm d}q^j
\label{afzct}
\end{equation}
satisfies
\begin{equation}
{\rm d}\theta=\omega.
\label{tgkmjnmm}
\end{equation}
We will also need the integral invariant of Poincar\'e--Cartan, denoted $\lambda$. If ${\cal H}$ denotes the Hamiltonian function, then $\lambda$ is 
defined as \cite{MS}
\begin{equation}
\lambda:=\theta+{\cal H}{\rm d}t.
\label{komome}
\end{equation}
Then the action (\ref{rmlcrlzl}) equals (minus) the line integral of $\lambda$,
\begin{equation}
S=-\int_{\mathbb{I}}\lambda.
\label{swws}
\end{equation}
On constant--energy submanifolds of $\mathbb{P}$, or else for fixed values of the time, we have
\begin{equation}
{\rm d}\lambda=\omega, \qquad {\cal H}={\rm const.}
\label{kuadraos}
\end{equation}

In what follows it will be convenient to drop the index $j$ while maintaining the index $\alpha$ of \v Cech cohomology. Let any three points $(q_{\alpha_1},p_{\alpha_1})$, $(q_{\alpha_2},p_{\alpha_2})$, $(q_{\alpha_3},p_{\alpha_3})$ be given on $\mathbb{P}$, respectively covered by coordinate charts $U_{\alpha_1}$, $U_{\alpha_2}$ and $U_{\alpha_3}$. Assume that $U_{\alpha_1}\cap U_{\alpha_2}\cap U_{\alpha_3}$ is nonempty, {\it i.e.},
\begin{equation}
U_{\alpha_1\alpha_2\alpha_3}:=U_{\alpha_1}\cap U_{\alpha_2}\cap U_{\alpha_3}\neq\phi,
\label{mnafd}
\end{equation}
and let $(q_{\alpha_{123}},p_{\alpha_{123}})$ be a variable point in this triple overlap,
\begin{equation}
(q_{\alpha_{123}},p_{\alpha_{123}})\in U_{\alpha_1\alpha_2\alpha_3}.
\label{ddmmd}
\end{equation}
{}Furthermore let $\mathbb{L}_{\alpha_1\alpha_2\alpha_3}(\alpha_{123})$ be a closed loop within $\mathbb{P}$ as constructed in ref. \cite{PHASEGERBES},
\begin{equation}
\mathbb{L}_{\alpha_1\alpha_2\alpha_3}(\alpha_{123}):=\mathbb{L}_{\alpha_1\alpha_2}(\alpha_{123})+\mathbb{L}_{\alpha_2\alpha_3}(\alpha_{123})+\mathbb{L}_{\alpha_3\alpha_1}(\alpha_{123}),
\label{bacon}
\end{equation}
where have explicitly indicated the dependence of the trajectory on the variable midpoint $(q_{\alpha_{123}},p_{\alpha_{123}})\in U_{\alpha_1\alpha_2\alpha_3}$. 
Altogether, the latter is traversed three times: once along the leg $\mathbb{L}_{\alpha_1\alpha_2}$ from $\alpha_1$ to $\alpha_2$,  once more along the leg $\mathbb{L}_{\alpha_2\alpha_3}$ from $\alpha_2$ to $\alpha_3$,  and finally along the leg $\mathbb{L}_{\alpha_3\alpha_1}$ from $\alpha_3$ to $\alpha_1$. For ease of writing, however, we will drop $\alpha_{123}$ from our notation. 

In the stationary--phase approximation, the 2--cocycle $g_{\alpha_1\alpha_2\alpha_3}^{(0)}$ defining a U(1) gerbe on $\mathbb{P}$ turns out to be \cite{PHASEGERBES}
\begin{equation}
g_{\alpha_1\alpha_2\alpha_3}^{(0)}=\exp\left(-\frac{{\rm i}}{\hbar}\int_{\mathbb{L}_{\alpha_1\alpha_2\alpha_3}^{(0)}}\lambda\right),
\label{yya}
\end{equation}
the superindex ${}^{(0)}$ standing for {\it evaluation at the extremal}, that is, at that closed loop $\mathbb{L}_{\alpha_1\alpha_2\alpha_3}^{(0)}$ of the type (\ref{bacon}) that renders the integral of $\lambda$ extremal. Equivalently, we can express $g_{\alpha_1\alpha_2\alpha_3}^{(0)}$ in terms of an integral over an extremal surface,
\begin{equation}
g_{\alpha_1\alpha_2\alpha_3}^{(0)}=\exp\left(-\frac{{\rm i}}{\hbar}\int_{\mathbb{S}_{\alpha_1\alpha_2\alpha_3}^{(0)}}\omega\right).
\label{axel}
\end{equation}
where $\mathbb{S}_{\alpha_1\alpha_2\alpha_3}^{(0)}$ is any surface bounded by the loop (\ref{bacon}). The cocycle is well defined in the sense that it does not depend on any {\it a priori}\/ choice of the points $\alpha_1$, $\alpha_2$ and $\alpha_3$. 

Eqn. (\ref{yya}) and its equivalent (\ref{axel}) give the stationary--phase approximation $g_{\alpha_1\alpha_2\alpha_3}^{(0)}$ to the 2--cocycle $g_{\alpha_1\alpha_2\alpha_3}$. The latter is a function of the variable midpoint (\ref{ddmmd}) through the extremal integration path $\mathbb{L}_{\alpha_1\alpha_2\alpha_3}^{(0)}$ or its equivalent extremal integration surface $\mathbb{S}_{\alpha_1\alpha_2\alpha_3}^{(0)}$, even if we no longer indicate this explicitly. Henceforth we will also drop the superindex $^{(0)}$, with the understanding that we are always working in the stationary--phase approximation. The latter is equivalent to the quantum--mechanical WKB approximation. Its role is that of minimising the symplectic area of the surface $\mathbb{S}_{\alpha_1\alpha_2\alpha_3}$. Now, in the WKB approximation, the absolute value of $\int_\mathbb{S}\omega/\hbar$ is proportional to the number of quantum--mechanical states contributed by the surface $\mathbb{S}$ \cite{LANDAU}. Hence the stationary--phase approximation applied here picks out those surfaces that contribute the least number of quantum--mechanical states. Moreover, since we are considering constant--energy surfaces $\mathbb{S}$, those states are stationary.

Concerning the connection on the gerbe \cite{GERBES}, one finds for the 1--form $A$  \cite{PHASEGERBES}
\begin{equation}
A=-\frac{{\rm i}}{\hbar}\lambda.
\label{ttwers}
\end{equation}
{}For the 2--form $B$ one finds, on constant--energy submanifolds of phase space,
\begin{equation}
B_{\alpha_2}-B_{\alpha_1}=-\frac{{\rm i}}{\hbar}\omega_{\alpha_1\alpha_2}.
\label{arfzktft}
\end{equation}
The above equation is interpreted as follows. Given the coordinate patches $U_{\alpha_1}$ and $U_{\alpha_2}$ such that $U_{\alpha_1}\cap U_{\alpha_2}$ is nonempty, let $\omega_{\alpha_1\alpha_2}$ denote the restriction of $\omega$ to $U_{\alpha_1}\cap U_{\alpha_2}$. Then a knowledge of $B$ on the patch $U_{\alpha_1}$ gives us the value of $B$ on the patch $U_{\alpha_2}$. Finally we have the 3--form
\begin{equation}
H={\rm d}B.
\label{kuadrados}
\end{equation}

\section{A U(1) invariance}\label{uuno}

By eqn. (\ref{swws}) we can perform the transformation
\begin{equation}
\lambda\longrightarrow\lambda+{\rm d}f,\qquad f\in C^{\infty}(\mathbb{P}),
\label{bbmj}
\end{equation}
where $f$ is an arbitrary function on $\mathbb{P}$ with the dimensions of an action, without altering the classical mechanics defined by $\omega$. Since the classical action $S$ is given by the line integral (\ref{swws}), the transformation (\ref{bbmj}) amounts to shifting $S$ by a constant $C$,
\begin{equation}
S\longrightarrow S+C,\qquad C:=-\int_{\mathbb{I}}{\rm d}f.
\label{chif}
\end{equation}
The way the transformation (\ref{bbmj}) acts on the quantum theory is well known. In the WKB approximation, the wavefunction reads \cite{LANDAU}
\begin{equation}
\psi_{\rm WKB}=R\exp\left(\frac{\rm i}{\hbar}S\right)
\label{ktfyrmlldmrd}
\end{equation}
for some amplitude $R$. Thus the transformation (\ref{bbmj}) multiplies the WKB wavefunction $\psi _{\rm WKB}$ and, more generally, any wavefunction $\psi$, by the {\it constant}\/ phase factor $\exp\left({\rm i}{C}/{\hbar}\right)$:
\begin{equation}
\psi\longrightarrow \exp\left(\frac{{\rm i}}{{\hbar}}{C}\right)\psi.
\label{llkbkb}
\end{equation}
Gauging the rigid symmetry (\ref{llkbkb}) one obtains the transformation law 
\begin{equation}
\psi\longrightarrow\Psi_f:= \exp\left(-\frac{{\rm i}}{{\hbar}}f\right)\psi, \qquad f\in C^{\infty}(\mathbb{P}),
\label{llmerk}
\end{equation}
$f$ being an arbitrary function on phase space, with the dimensions of an action.  Now eqn. (\ref{llmerk}) implies that, if the original wavefunction $\psi$ depends only on the coordinates $q$, its transform $\Psi_f$ under an arbitrary $f\in C^{\infty}(\mathbb{P})$ generally depends also on the momenta $p$. According to standard lore this is prohibited by Heisenberg's uncertainty principle.  Moreover, even if wavefunctions can be defined on phase space, the local transformations (\ref{llmerk}) need not be a symmetry of our theory. We address these two points separately in sections \ref{wwff} and  \ref{unosymp}.

\section{Probability distributions on phase space}\label{wwff}

Concerning the first objection raised above one should observe that phase--space quantum mechanics, {\it while respecting the constraints imposed by Heisenberg's principle}, is almost as old as quantum mechanics itself \cite{WIGNER}; we refer the reader to \cite{GOSSON, ZACHOS} for a compilation of relevant literature. We will henceforth  call the objects $\Psi_f=\Psi_f(q,p)$ introduced in (\ref{llmerk})  {\it probability distributions}; they are defined on $\mathbb{P}$.  For simplicity, in what follows we will omit the subscript ${}_f$ from $\Psi_f$.

Specifically, in refs. \cite{LETTER, MAURICE} it has been shown that  the usual Schr\"odinger equation for the usual wavefunction 
$\psi=\psi(q)$,
\begin{equation}
{\cal H}\left(q,-{\rm i}\hbar{\partial_q}\right)\psi(q)=E\psi(q),
\label{neu}
\end{equation}
is equivalent  to the following Schr\"odinger--like equation for the probability distribution $\Psi=\Psi(q,p)$ on $\mathbb{P}$:
\begin{equation}
{\cal H}\left(\frac{q}{2}+{\rm i}\hbar{\partial_p}, \frac{p}{2}-{\rm i}\hbar{\partial_q}\right)\Psi(q,p)=E\Psi(q,p).
\label{lvzmsrb}
\end{equation}
Moreover, the quantum operators
\begin{equation}
Q_{A_0'}:=\frac{q}{2}+{\rm i}\hbar{\partial_p}, \qquad P_{A_0'}:=\frac{p}{2}-{\rm i}\hbar{\partial_q}
\label{lineop}
\end{equation}
satisfy the usual canonical commutation relations
\begin{equation}
[Q_{A_0'}, P_{A_0'}]={\rm i}\hbar,
\label{werg}
\end{equation}
so eqn. (\ref{lvzmsrb}) can be rewritten as
\begin{equation}
{\cal H}\left(Q_{A_0'}, P_{A_0'}\right)\Psi(q,p)=E\Psi(q,p).
\label{vzzhjp}
\end{equation}
A computation shows that $\Psi(q,p)$ in (\ref{lvzmsrb}) and $\psi(q)$ in (\ref{neu}) are related as per eqn. (\ref{llmerk}), the argument $f(q,p)$ of this latter exponential being
\begin{equation}
f_{A_0'}(q,p):=\frac{1}{2}pq=\frac{1}{2}p_jq^j .
\label{maurix}
\end{equation}
That is, the Schr\"odinger eqns. (\ref{neu}) and (\ref{lvzmsrb}) are equivalent if, and only if,  the respective probability amplitude $\Psi(q,p)$ and wavefunction $\psi(q)$ are related as
\begin{equation} 
\Psi(q,p)=\exp\left(-\frac{{\rm i}}{2{\hbar}}pq\right)\psi(q).
\label{laer}
\end{equation}
Eqn. (\ref{laer}) is in perfect agreement with the results of refs. \cite{LETTER, MAURICE}. 

The reason for the subindex $A_0'$ in (\ref{lineop})--(\ref{maurix}) above is the following. Consider the {\it symplectic}\/ exterior derivative on phase space,
\begin{equation}
{\rm d}':=-{\rm d}q\,\partial_q+{\rm d}p\,\partial_p.
\label{ableitung}
\end{equation}
Consider also the following connection $A_0'$ on phase space:
\begin{equation}
A_0':=-\frac{{\rm i}}{\hbar}{\rm d}f_{A_0'}=\frac{1}{2{\rm i}\hbar}\left(p\,{\rm d}q+q\,{\rm d}p\right).
\label{goss}
\end{equation}
Let us now covariantise ${\rm d}'$ as
\begin{equation}
{\rm d}'\longrightarrow D_{A_0'}':={\rm d}'+ A_0'.
\label{pptrmlla}
\end{equation}
We see that the operators of  eqn. (\ref{lineop}) are the result of gauging the symplectic derivative ${\rm d}'$  by the connection $ A_0'$:
\begin{equation}
{\rm i}\hbar D_{A_0'}'={\rm d}q\left(\frac{p}{2}-{\rm i}\hbar\partial_q\right)+{\rm d}p\left(\frac{q}{2}+{\rm i}\hbar\partial_p\right).
\label{platsch}
\end{equation}
Covariantising the symplectic derivative as per eqn. (\ref{platsch}) is equivalent to the symplectic transformation considered in refs. \cite{LETTER, MAURICE} that renders the quantum theory manifestly symmetric under the symplectic exchange of $q$ and $p$. This latter symmetry is conspicuously absent in the usual formulation of quantum mechanics based on the usual Schr\"odinger equation (\ref{neu}).  

One can consider more general covariantisations of the symplectic derivative (\ref{ableitung}). Given a solution $\psi=\psi(q)$ of the usual Schr\"odinger equation (\ref{neu}), and given a function $f_{A'}\in C^{\infty}(\mathbb{P})$, define $\Psi=\Psi(q,p)$ as per eqn. (\ref{llmerk}). We can require the latter to satisfy a phase--space Schr\"odinger equation, that we can determine as follows. One picks a certain connection
\begin{equation}
A'=\frac{1}{{\rm i}\hbar}\left[A'_q(q,p){\rm d}q+A'_p(q,p){\rm d}p\right]
\label{nocc}
\end{equation}
that one takes to covariantise the symplectic derivative ${\rm d}'$ of (\ref{ableitung}),
\begin{equation}
D_{A'}':={\rm d}'+A'.
\label{vacovv}
\end{equation}
The components $A_q'=A'_q(q,p)$ and $A_p'=A'_p(q,p)$ are unknown functions of $q,p$. However they are not totally unconstrained, because the position and momentum operators 
\begin{equation}
Q_{A'}:=A_p'+{\rm i}\hbar\partial_p, \qquad P_{A'}:=A_q'-{\rm i}\hbar\partial_q
\label{edfre}
\end{equation}
will enter the Hamiltonian ${\cal H}(Q_{A'}, P_{A'})$ obtained from ${\cal H}(Q=q, P=-{\rm i}\hbar\partial_q)$ by the replacements $Q\rightarrow Q_{A'}$, $P\rightarrow P_{A'}$: 
\begin{equation}
{\cal H}\left(Q_{A'}, P_{A'}\right)=\frac{1}{2m}P_{A'}^2+V(Q_{A'})=\frac{1}{2m}\left(A_q'-{\rm i}\hbar\partial_q\right)^2+V(A_p'+{\rm i}\hbar\partial_p).
\label{tse}
\end{equation}
As such, the operators (\ref{edfre}) must satisfy the canonical commutation relations (\ref{werg}). This requires that the following {\it integrability condition}\/ hold:
\begin{equation}
\frac{\partial A_p'}{\partial q}+\frac{\partial A_q'}{\partial p}=1.
\label{frz}
\end{equation}
Notice the positive sign, instead of negative, between the two summands on the left--hand side of (\ref{frz}). This is ultimately due to the fact that we are covariantising the symplectic derivative ${\rm d}'$ rather than the usual exterior derivative ${\rm d}={\rm d}q\,\partial_q+{\rm d}p\,\partial_p$. A computation shows that the phase--space Schr\"odinger equation
\begin{equation}
{\cal H}(Q_{A'}, P_{A'})\Psi(q,p)=E\Psi(q,p)
\label{slke}
\end{equation}
is equivalent to the usual Schr\"odinger equation (\ref{neu}) if, and only if, $A_q'$, $A_p'$ and $f_{A'}$ are related as
\begin{equation}
A_q'=\partial_q f_{A'},\qquad A_p'=q-\partial_pf_{A'}.
\label{ttraex}
\end{equation}
When eqn. (\ref{ttraex}) holds,  the integrability condition (\ref{frz}) is automatically satisfied. We conclude that picking one $f_{A'}\in C^{\infty}(\mathbb{P})$ and defining the connection $A'$ as per eqns. (\ref{nocc}), (\ref{ttraex}), we arrive at the phase--space wave equation (\ref{slke}). Alternatively, given a connection (\ref{nocc}) and a phase--space wave equation (\ref{slke}), we can find a function $f_{A'}\in C^{\infty}(\mathbb{P})$, defined by (\ref{ttraex}) up to integration constants, such that the corresponding probability distribution $\Psi(q,p)$ is related to the wavefunction $\psi(q)$ as per eqn. (\ref{llmerk}), where $f=f_{A'}$. Eqn. (\ref{ttraex}) above gives us a whole $C^{\infty}(\mathbb{P})$'s worth of phase--space Schr\"odinger equations, one per each choice of a function $f_{A'}$. The latter may well be termed the {\it generating function}\/ for the transformation  (\ref{llmerk}) between configuration--space and phase--space probability distributions and their corresponding Schr\"odinger equations.

Given a connection $A'$ as per eqns. (\ref{nocc}) and (\ref{ttraex}), how is $A'$ is related to the potential 1--form $A$ on the gerbe, eqn. (\ref{ttwers})? The answer to this question will be given in section \ref{unosymp}; it necessitates the notion of gauge transformations on the gerbe, which we introduce next.

\section{Gauge transformations}\label{bbchpebb}

Given an arbitrary function $f\in C^{\infty}(\mathbb{P})$, the triple of forms $A, B, H$ on the gerbe transform under the local U(1) group of eqn. (\ref{llmerk}) as
\begin{equation}
\delta_0 A:=-\frac{{\rm i}}{\hbar}{\rm d}f,\qquad\delta_0 B=0, \qquad \delta_0 H=0,\qquad f\in C^{\infty}(\mathbb{P}).
\label{hhyy}
\end{equation}
The gauge transformations eqn. (\ref{hhyy}) are formally identical to the U(1) gauge transformations of electromagnetism. There are, however, three key differences:\\
{\it i)} the Noether charge of electromagnetism may, but need not, be present here. Should electric charges $e$ exist, one could introduce an {\it electromagnetic}\/ potential $A_e$ and its corresponding field--strength $F_e:={\rm d}A_e$. This however would be an additional U(1) symmetry, implemented by a fibre bundle instead of a gerbe;\\
{\it ii)} the covariant derivative of electromagnetism is ${\rm d}+eA_e$, while that considered here is ${\rm d}'+A'$;\\
{\it iii)} the 2--form d$A$ on phase space is not a field strength but the defining equation of the Neveu--Schwarz 2--form potential $B$.\\
Altogether we conclude that $A$ is not an electromagnetic potential, nor is the corresponding U(1) that of electromagnetic gauge invariance.

The gauge transformations (\ref{hhyy}) by no means exhaust all possibilities for U(1) transforming the connection on the gerbe.  On phase space let us consider an arbitrary 1--form $\varphi\in\Omega^1(\mathbb{P})$ with the dimensions of an action.  We define a second set of U(1) gauge transformations:
\begin{equation}
\delta_1 A:=-\frac{{\rm i}}{\hbar}\varphi, \qquad\delta_1 B=-\frac{{\rm i}}{\hbar}{\rm d}\varphi,\qquad\delta_1 H=0, \qquad \varphi\in\Omega^1(\mathbb{P}).
\label{rmllcmm}
\end{equation}
We observe that $\delta_1$ is parametrised by a 1--form $\varphi$ while $\delta_0$ had a 0--form $f_{A'}$  as its gauge parameter. The $\delta_1$ gauge transformation law of the wavefunction is
\begin{equation}
\psi\longrightarrow\Psi_{\varphi}:=\exp\left(-\frac{{\rm i}}{\hbar}\varphi\right)\psi,Ê\qquad \varphi\in\Omega^1(\mathbb{P}).
\label{rmlldftcmm}
\end{equation}
After this transformation, the probability distribution $\Psi_{\varphi}$ is no longer a function, but a nonhomogeneous differential form on phase space. We will analyse this important fact in a forthcoming paper \cite{NOI}, where the link between our approach and that of ref. \cite{MATONE} will also be examined.

\section{U(1) gauge invariance and symplectic covariance}\label{unosymp}

We can now answer the question posed at the end of section \ref{wwff}, namely: given a connection $A'$ as per eqns. (\ref{nocc}) and (\ref{ttraex}), can one  
$\delta_0$-- and/or $\delta_1$--transform the potential 1--form $A$ on the gerbe so that $A'=A+\delta A$? That is, can $A'$ and $A$ be gauge equivalent?

Consider $\delta_1$--transformations first. We are looking for a 1--form $\varphi=\varphi_q{\rm d}q+\varphi_p{\rm d}p$ such that $A+\delta_1A=A+\varphi/({\rm i}\hbar)$ will equal the given $A'$ of eqns. (\ref{nocc}) and (\ref{ttraex}). One immediately verifies that 
\begin{equation}
\varphi_q(q,p):=p+\partial_qf_{A'}, \qquad \varphi_p(q,p):=q-\partial_pf_{A'}
\label{fwpoouh}
\end{equation}
meets our requirements, hence any $A'$ is $\delta_1$--gauge equivalent to the potential 1--form $A$ on the gerbe.  

However, $\delta_0$--gauge transformations are more restrictive. In this case we have to set $\varphi_q=\partial_qF(q,p)$ and $\varphi_p=\partial_pF(q,p)$ for a certain function $F\in C^{\infty}(\mathbb{P})$. The latter is to be determined by integration of the system of equations
\begin{equation}
\partial_qF=p+\partial_qf_{A'},  \qquad \partial_pF=q-\partial_pf_{A'},
\label{msssy}
\end{equation}
for a given generating function $f_{A'}\in C^{\infty}(\mathbb{P})$. A solution to (\ref{msssy}) can exist only when 
\begin{equation}
\partial_{q^j}\partial_{p_k}f_{A'}=0, \qquad \forall j,k=1,\ldots d.
\label{lapla}
\end{equation}
The general solution to (\ref{lapla}) is the sum of a function of coordinates only and a function of momenta only,
\begin{equation}
f_{A'}(q,p)=g(q)+h(p).
\label{dalembert}
\end{equation}
So only when the generating function $f_{A'}(q,p)$ of the given connection $A'$ satisfies condition (\ref{dalembert}) can one find a $\delta_0$--gauge transformation that will render $A'$ gauge equivalent to the potential 1--form $A$ on the gerbe (\ref{ttwers}).

This brings us back to the second objection raised after eqn. (\ref{llmerk}), that we can finally answer in the affirmative. The local transformations (\ref{llmerk}) {\it are}\/ a symmetry of our theory, in the sense already explained in section \ref{wwff}. Namely, the transformation (\ref{llmerk}) from $\psi(q)$ to $\Psi(q,p)$ must be accompanied by the corresponding covariantisation (\ref{vacovv}) of the symplectic derivative ${\rm d}'$ within the Schr\"odinger equation. Since the connection $A'$ and the potential 1--form $A$ on the gerbe are gauge equivalent (this is always the case under $\delta_1$, and also under $\delta_0$ whenever condition (\ref{dalembert}) holds), this can be understood as a covariantisation of the symplectic derivative ${\rm d}'$ within the Hamiltonian operator, by means of the potential 1--form $A$ on the gerbe. Therefore we replace eqn. (\ref{vacovv}) with the following covariant derivative:
\begin{equation}
D'_A:={\rm d}'+A,
\label{vvaccb}
\end{equation}
where $A$ is the potential 1--form on the gerbe. So we can always covariantise the symplectic derivative ${\rm d}'$ as per eqn. (\ref{vvaccb}) thanks to the existence of a gerbe on classical phase space.

To summarise, {\it gauging the rigid symmetry (\ref{llkbkb}), {\it i.e.}, allowing for the local transformations (\ref{llmerk}), one arrives naturally at a phase--space formulation of quantum mechanics}. In other words, {\it U(1) gauge invariance on the gerbe is equivalent to symplectic covariance}, the latter understood as in refs. \cite{LETTER, MAURICE}:  as the possibility to U(1)--rotate the Schr\"odinger equation from configuration space into phase space, and also within the latter itself, with a point--dependent rotation parameter.
\vskip2cm
\noindent
{\bf Acknowledgements} Both authors would like to thank Albert--Einstein--Institut (Potsdam, Germany) for hospitality during the preparation of this article and Profs. H. Nicolai and S. Theisen for their kind invitation. This work has been supported by Ministerio de Educaci\'{o}n y Ciencia (Spain) through grant FIS2005--02761, by Generalitat Valenciana, by EU FEDER funds, by EU network MRTN--CT--2004--005104 ({\it Constituents, Fundamental Forces and Symmetries of the Universe}), and by Deutsche Forschungsgemeinschaft.

\end{document}